\documentclass{aa}
\usepackage{graphics,latexsym,epsfig,graphicx,amssymb,natbib}     

\bibpunct{(}{)}{;}{a}{}{,}
\renewcommand{\d}[0]{{\rm d}}

\newcommand{\ave}[1]{\langle #1 \rangle}

\newcommand{\Ref}[1]{(\ref{#1})}

\newcommand{\changed}[1]{#1}

\begin{document}

\title{How accurate is Limber's equation?}
\author{P. Simon}
\institute{%
  Institute for Astronomy, University of Edinburgh, Royal
  Observatory, Blackford Hill, Edinburgh EH9-3HJ, UK\\
  \email{psimon@roe.ac.uk}\\
  Argelander-Institut f\"ur
  Astronomie\thanks{Founded by merging of the Institut f\"ur
    Astrophysik und Extraterrestrische Forschung, the Sternwarte, and
    the Radioastronomisches Institut der Universit\"at Bonn.}, Universit\"at Bonn,
  Auf dem H\"ugel 71, 53121 Bonn, Germany}

\date{Received/Accepted} \authorrunning{P. Simon} \titlerunning{}

\keywords{}
\abstract{}
{The so-called Limber equation is widely used
 in the literature to relate the projected angular clustering of
 galaxies to the spatial clustering of galaxies in an approximate
 way. This note gives estimates of where the regime of applicability
 of Limber's equation stops.}
{This paper revisits Limber's equation, summarises its underlying
  assumptions, and compares its predictions to the accurate integral
  of which the Limber equation is an approximation for some realistic
  situations. Thereby, the accuracy of the approximation is
  quantified. Different cases of spatial correlation functions are
  considered.}
{Limber's equation is accurate for small galaxy separations but breaks
  down beyond a certain separation that depends mainly on the ratio
  \mbox{$\sigma/r_{\rm m}$} and to some degree on the power-law
  index of spatial clustering; $\sigma$ is the $1\sigma$-width of the
  galaxy distribution in \emph{comoving distance}, and $r_{\rm m}$ the
  mean comoving distance. Limber's equation introduces a $10\%$
  systematic error that may be important at a scale of about a few
  degrees or in some cases, narrow galaxy distributions for example,
  even at arcmin scales. As rule-of-thumb, a $10\%$ relative error is
  reached at \mbox{$\sim260\,\sigma/r_{\rm m}$ arcmin} for
  $\gamma\sim1.6$, if the spatial clustering is a power-law
  \mbox{$\xi\propto r^{-\gamma}$}. More realistic $\xi$ are discussed
  in the paper.}
{Limber's equation becomes increasingly inaccurate for larger angular
separations. Ignoring this effect and blindly applying Limber's
equation can possibly bias results for the inferred spatial
correlation.  It is suggested to use in cases of doubt, or maybe even
in general, the exact equation that can easily be integrated
numerically in the form given in the paper. This would actually
eliminate all problems discussed in this paper.}

\keywords{galaxies - statistics : cosmology - large-scale structure of
  Universe : cosmology - theory}

\maketitle

\section{Introduction}

If the distance of individual galaxies is not known, but only
statistically for a sample of galaxies on the whole, one has to infer
the real-space correlation from the observed angular correlation on
the sky. This task has been performed many times since the early days
of galaxy surveys as, for example, in the pioneering work of
\citet{tot69}, \citet{groth77} and \citet{davis77} based on the Lick
galaxy survey. To relate the projected correlation function,
$\omega(\theta)$, to the spatial, three-dimensional correlation
function, $\xi$, many authors use an approximation that was introduced
by \citet{limber53}. This approximation is nowadays referred to as
Limber's equation.

It will be shown in Sect. \ref{limberapprox} that Limber's
approximation diverges when galaxies are closely distributed.
That Limber's equation breaks down for narrow distributions is no
surprise, as it is assumed that the distributions do no vary
considerably over the coherence length of the spatial correlation
function. But how wide has a distribution to be?  How much do we bias
our estimate for the angular correlation by using Limber's equation,
and is there an alternative?  These questions are addressed by this
paper.

\citet[ p.199]{peebles80} argued that this break-down occurs at
separations of roughly $30^\circ$, but made assumptions about galaxy
surveys that do not meet the standards of many contemporary wide-field
surveys. Due to the availability of spectroscopic or photometric
redshifts quite narrowly binned distributions are often used in
current studies.

This paper delineates the circumstances in which Limber's
approximation is valid and gives an alternative for situations where
it is not valid. Throughout this paper a flat universe will be
assumed.

The exact equation for $\omega$ is given in Sect.
\ref{exactsolution}. The thereafter following Sect.
\ref{limberapprox} summarises Limber's approximation and investigates
why it breaks down for peaked galaxy distributions or beyond a certain
angular separation. Alternative approximations for cases of narrow
distributions of galaxies or for larger $\theta$ are given inside the
Sect.  \ref{thinlayer}.  By looking at different distributions with
varying widths and means, the accuracy of the Limber approximation is
assessed in Sect.  \ref{accuracy}. The paper finishes with conclusions
in Sect.  \ref{conclusions}.

\section{Relation between spatial and angular correlation function}\label{defsect}

\subsection{Exact relation with only some restrictions}\label{exactsolution}

Consider two number density fields of galaxies, $n_1(\vec{r})$ and
$n_2(\vec{r})$, as a function of the position $\vec{r}$ within the
comoving frame. One defines the spatial clustering, or spatial
correlation function, $\xi(\Delta\chi)$ in terms of the average number
of pairs of $n_1$- and $n_2$-galaxies that can be found when
considering small volumes $\d V_{1/2}$ with a separation of
$\Delta\chi=|\vec{r}_1-\vec{r}_2|$:
\begin{equation}
  \ave{n_1(\vec{r}_1)n_2(\vec{r}_2)}\d V_1\d V_2=
  \ave{n_1}\ave{n_2}[1+\xi(\Delta\chi)]\d V_1\d V_2\;.
\end{equation}
$\ave{.}$ denotes the ensemble average.  It is already assumed here
that the random fields $n_{1,2}$ are statistically homogeneous and
isotropic so that $\xi$ is a function of the distance $\Delta\chi$ only.
If we have $n_1=n_2$, then $\xi(\Delta\chi)$ is a auto-correlation function of
a galaxy number density field. 

Observing the galaxy density fields as projections on the sky,
\begin{eqnarray}\label{projectionofn}
  \hat{n}_i(\vec{\theta})&=&
  \int_0^\infty\,\d
  r\,q_i(r)\,n_i\left(r\vec{\theta},r\right)\;,
\end{eqnarray}
yields a different correlation function, $\omega(\theta)$, when pairs
of galaxies, $\hat{n}_1$ and $\hat{n}_2$, within solid angles
$\d\Omega_{1/2}$ of angular separation $\theta$ are considered:
\begin{equation}
  \ave{\hat{n}_1(\vec{\theta}_1)\hat{n}_2(\vec{\theta}_2)}\d\Omega_1\d\Omega_2=
  \ave{\hat{n}_1}\ave{\hat{n}_2}[1+\omega(\theta)]\d\Omega_1\d\Omega_2\;.
\end{equation}
Here, $q_i(r)$ is a filter or weight function (of the number density)
solely depending on the comoving radial distance $r$. In the context
of galaxy number densities, this filter needs to be normalised to one,
i.e.
\begin{equation}
\int_0^\infty\d r\,q_i(r)=1\;;
\end{equation}
it specifies the frequency of galaxies within the distance interval
\mbox{$[r,r+\d r[$}. \changed{The details of the filter depend on the
selection function of the survey.} The vector $\vec{\theta}$ is the
direction of the line-of-sight.  The observer is sitting at the origin
of the reference frame, \mbox{$\vec{r}=0$}, while other points on the
line-of-sight are parametrised by $r\vec{\theta}$. To take into
account a possible time-evolution of $n_i$ the second argument in
$n_i(r\vec{\theta},r)$ is supposed to parametrise the look-back time,
$t(r)$, as a function of the comoving radial distance at which the
random field is observed:
\begin{equation}
  t(r)=\frac{1}{c}\int_0^r\d r^\prime\,a(r^\prime)\;,
\end{equation}
where $a(r)$ is the scale factor at comoving distance $r$, and $c$ the
vacuum speed of light.

As already mentioned the arguments stated above assume a Euclidean
geometry for the Universe. This is not a strong restriction, though,
because any other fiducial cosmological model can be incorporated by
redefining the filter $q_i$ -- by absorbing $F(x)$ in the
(relativistic) Limber equation of \citet{peebles80} (Chapter III,
Sect. 56) into $q_i$ -- which leaves the form of \Ref{projectionofn}
effectively unchanged.

Let us now derive a relation between $\omega$ and $\xi$.  Imagine we
observe the spatial fluctuations (\mbox{$\ave{n_i}(r)$} is the mean density at cosmic
time $t(r)$),
\begin{equation}\label{dencontrast}
  \delta_{1,2}(r\vec{\theta},r)\equiv
  \frac{n_{1,2}(r\vec{\theta},r)}{\ave{n_{1,2}}(r)}-1\;,
\end{equation}
of the two random fields as projection on the sky towards the
direction $\vec{\theta}$:
\begin{equation}\label{projectionofd}
  \hat{\delta}_{1,2}(\vec{\theta})=
  \int_0^\infty\d r\,p_{1,2}(r)\delta_{1,2}(r\vec{\theta},r)\;.
\end{equation}
Here, $p_i$ are the filters of the spatial fluctuations, $\delta_i$,
that correspond to $q_i$ (the filter of the density
fields):\footnote{In this paper the focus will be on the angular
clustering and hence normalised $p_i$ are assumed, albeit all
equations and the conclusions are also valid for other types of
projections.}
\begin{equation}
  p_i(r)=q_i(r)\,\ave{n_i}(r)\left[\int_0^\infty\d r\,q_i(r)\,\ave{n_i}(r)\right]^{-1}\;.
\end{equation}
This relation between $q_i$ and $p_i$ follows from the definition of
the density contrast of the projected (number) density:
\begin{equation}
  \hat{\delta}_i(\vec{\theta})=\frac{\hat{n}_i(\vec{\theta})}{\ave{\hat{n}_i}}-1\;.
\end{equation}
If the mean density is constant, both associated filters $p_i$ and
$q_i$ will be identical.

According to the definition of $\omega$ we have for sight-lines
$\vec{\vartheta}_1$ and $\vec{\vartheta}_2$ spanning an angle
\mbox{$\theta=\sphericalangle\vec{\vartheta}_1,\vec{\vartheta}_2$}:
\begin{eqnarray}
  &&\omega(\theta)=
  \ave{\hat{\delta_1}(\vec{\vartheta}_1)\hat{\delta}_2(\vec{\vartheta}_2)}\\
  \nonumber&&=
  \int_0^\infty\!\!\!\!\d r_1\int_0^\infty\!\!\!\!\d r_2\,p_1(r_1)p_2(r_2)
  \ave{\delta_1(r_1\vec{\vartheta}_1,r_1)\delta_2(r_2\vec{\vartheta}_2,r_2)}\\
  \label{laststep}
&&\approx\int_0^\infty\d r_1\int_0^\infty\d r_2\,p_1(r_1)p_2(r_2)\,
\xi\!\left(R,\frac{r_1+r_2}{2}\right)\;,
\end{eqnarray}
where 
\begin{equation}
  R\equiv\sqrt{r_1^2+r_2^2-2r_1r_2\cos{\theta}}\;.
\end{equation}
As before the second argument of the spatial correlation, $\xi(R,r)$,
parametrises the time, $t(r)$, at which the spatial correlation
function is observed.

Two assumptions had to be made to arrive at \Ref{laststep}: a) the
random fields $\delta_{1,2}$, and hence also their projections,
$\hat{\delta}_{1,2}$, are statistically isotropic and homogeneous, and
b) the time-evolution of $\xi$ is small within the region where the
product $p_1(r_1)p_2(r_2)$ is non-vanishing. Due to assumption a)
$\omega$ depends only on $\theta$ and is independent of the directions
$\vec{\vartheta}_{1,2}$. Owing to assumption b) we can approximate the
spatial correlation of fluctuations at different cosmic times (radial
distances) $t(r_1)$ and $t(r_2)$ by a representative $\xi$ at time
$(r_1+r_2)/2$.  

Note that Eq.  \Ref{laststep} is, under the previously stated
assumptions, valid even for large $\theta$.

\subsection{Limber's approximation and its breakdown}\label{limberapprox}

\changed{\citet{limber53} introduced several approximations in order
to simplify the integral Eq. \Ref{laststep}. The details of this
approximation are given in Appendix \ref{limberdetails}. Here only the
result is quoted: 
\begin{eqnarray}\label{limber}
  \omega(\theta)&=&
  \int_0^\infty\d\bar{r}\,p_1(\bar{r})p_2(\bar{r})
  \int_{-\infty}^{+\infty}\d{\Delta r}\,\,\xi(R,\bar{r})\\
  R&\equiv&\sqrt{\bar{r}^2\theta^2+\Delta r^2}\;.
\end{eqnarray}}

The useful Eq. \Ref{limber} has frequently been used in the astronomy
community because it allows one to find an analytical expression
\citep{peebles80} for $\omega$ in case of a power-law like
\begin{equation}\label{powerlawxi}
  \xi(r)=\left(\frac{r}{r_0}\right)^{-\gamma}\;,
\end{equation}
($r_0$ is the clustering length) namely
\begin{equation}\label{limberomega}
\omega(\theta)=A_\omega\left(\frac{\theta}{1\,\rm RAD}\right)^{1-\gamma}\;.
\end{equation}
The angular clustering amplitude at $\theta=1\,\rm RAD$ is
\begin{equation}\label{aomega}
  A_\omega=
  \sqrt{\pi}\,r_0^\gamma\frac{\Gamma(\gamma/2-1/2)}{\Gamma(\gamma/2)}\!\!
  \int_0^\infty\d\bar{r}\,p_1(\bar{r})p_2(\bar{r})\bar{r}^{1-\gamma}\;.
\end{equation}
It has been found that a power-law $\xi$ is a fairly good description
of the true spatial correlation function \citep[e.g. ][]{zehavi02} of
galaxies \changed{for scales below \mbox{$\sim 15\,\rm Mpc/h$}}, which
makes Limber's equation quite practical. Typical values are
\mbox{$\gamma\sim1.8$} and \mbox{$r_0\sim5h^{-1}\rm Mpc$}.
 
However, due to the integral, $A_\omega$ diverges to infinity if the
width of the distributions $p_{1,2}$ approaches zero and as long as
$p_{1,2}$ remain overlapping; the latter is trivially true if we are
considering auto-correlations, $p_1=p_2$. To make this point, let us
assume that $p_{1,2}$ are top-hat functions with centre $r_{\rm c}$
and width $2\Delta r$:
\begin{equation}\label{tophat}
  p_{1,2}(r)=
  \frac{1}{2\Delta r}\times
  \left\{
    \begin{array}{ll}
      1&{\rm for~} r\in[r_{\rm c}-\Delta r,r_{\rm c}+\Delta r]\\
      0&{\rm otherwise}
    \end{array}
  \right.\;.
\end{equation}
It follows for $A_\omega$:
\begin{equation}\label{approx}
  A_\omega\propto\frac{1}{4\Delta r^2}
  \int_{r_{\rm c}-\Delta r}^{r_{\rm c}+\Delta
    r}\d\bar{r}\,\bar{r}^{1-\gamma}
   \approx\frac{1}{2\Delta r}r_{\rm c}^{1-\gamma}\;.
\end{equation}
The last step is valid for small \mbox{$\Delta r\ll r_{\rm c}$}, which
shows that $A_\omega$ diverges for narrow distributions as $\Delta
r^{-1}$.  This small calculation implies that Limber's equation
possibly over-estimates the angular correlation $\omega$ to some
degree.

\changed{In the astronomical literature, Eq.  \Ref{limber} is also
known in other forms involving the three-dimensional power spectrum
\citep[e.g.][]{kaiser98,hamana04}. No matter the form of Limber's
equation, it always suffers from the divergence previously discussed
and from the inaccuracies that are going to be discussed in the
following.}

\subsection{Thin-layer approximation}\label{thinlayer}

Approximation Eq. \Ref{limber} fails if the weight functions become
delta-like functions, i.e. for galaxies being located inside one layer
at comoving distance $r_{\rm c}$. To obtain the correct solution for
\begin{equation}
  p_{1,2}(r)=\delta_{\rm D}(r-r_{1,2,\rm c})
\end{equation}\label{deltaomega1}
(where $\delta_{\rm D}(r)$ is the Dirac delta function) we simply have
to go back to Eq. \Ref{laststep} and find:
\begin{equation}
  \omega(\theta)=
  \xi\left(\sqrt{r_{1,\rm c}^2+r_{2,\rm c}^2-2r_{1,\rm c}r_{2,\rm
        c}\cos{\theta}},\frac{r_{1,\rm c}+r_{2,\rm c}}{2}\right)\;.
\end{equation}
For \mbox{$r_{1,\rm c}=r_{2,\rm c}\equiv r_{\rm c}$}, i.e. for
auto-correlations which are mainly considered in this paper, we obtain
as quite simple and intuitive result
\begin{equation}\label{deltaomega2}
  \omega(\theta)=
  \xi\!\left(r_{\rm c}\sqrt{2}\sqrt{1-\cos{\theta}},r_{\rm c}\right)
  \approx\xi(r_{\rm c}\theta,r_{\rm c})\;.
\end{equation}
This means that for very narrow filter we essentially observe a
rescaled $\xi$. An immediate consequence is that $\omega$ has
\emph{the same slope}, $\gamma$, as $\xi$, if $\xi$ happens to be a
power-law as in Eq. \Ref{powerlawxi}; on the other hand for wide
filter, in the regime where Limber's approximation is accurate,
$\omega$ has a shallower slope of $\gamma-1$.

\changed{It can be shown, as done in Appendix \ref{largertheta}, that
the thin-layer approximation is not just a sole particularity of
extremely peaked weight functions $p$. The solution is asymptotically
approached by \Ref{laststep} if the separation $\theta$ gets large
enough, namely \mbox{$\theta\gtrsim\sqrt{2}\frac{\sigma}{r_{\rm m}}$}
in the case of auto-correlations $p(r)\equiv p_1(r)=p_2(r)$; $\sigma$
is the $1\sigma$-width and $r_{\rm m}$ the mean of the weight $p(r)$
(both in units of comoving distance).}

\changed{This discussion shows that Limber's equation and the
thin-layer equation are approximations for two extremes, namely for
small $\theta$ and large $\theta$, respectively. The transition point
between the two regimes is mainly determined by the ratio
$\sigma/r_{\rm m}$ implying that the accuracy of Limber's equation is,
in the case of auto-correlations at least, mainly determined by this
ratio.}

\subsection{Numerical integration of the exact equation}

\changed{With the availability of computers there is no longer a need
to use any approximations in this context.}  However, one finds that
solving Eq. \Ref{laststep} (or \ref{exactsol}) by numerical
integration with a power-law $\xi$ is cumbersome because many fine
bins are needed to achieve the desired accuracy. This is mainly due to
the argument $R$ in $\xi$ which does not sample $\xi$ on a log-scale,
which however should be aimed at with a power-law like $\xi$. It is
advisable to consider a numerical integration with different
coordinates such that the spatial part of $\xi$ depends just on one
integration variable $R$.

This can be done \changed{by changing the integration variable in
  the inner integral of \Ref{exactsol} from $\Delta r$ to $R$}:
\begin{equation}\label{omega0}
  \omega(\theta)=\frac{1}{1+\cos{\theta}}
  \int_0^\infty\!\!\!\d\bar{r}\!\!\!\!\!\!
  \int^{2\bar{r}}\limits_{\bar{r}\sqrt{2(1-\cos{\theta})}}\!\!\!\!\!\!\!\!\!\!\d R\,\,
  Q(\bar{r},\Delta)\,\xi(R,\bar{r})\,\frac{R}{\Delta}\;,
\end{equation}
where
\begin{equation}
  \Delta\equiv\frac{1}{\sqrt{2}}\sqrt{\frac{R^2-2\bar{r}^2(1-\cos{\theta})}{1+\cos{\theta}}}\;, 
\end{equation}
and 
\begin{equation}
Q(\bar{r},\Delta)\equiv
p_1(\bar{r}-\Delta)p_2(\bar{r}+\Delta)+p_1(\bar{r}+\Delta)p_2(\bar{r}-\Delta)\;.
\end{equation}

The expression \Ref{omega0} may look more complicated than
\Ref{laststep} or \Ref{exactsol} but is superior for numerical
integrations: The integration over $R$ can be done on a logarithmic
scale, the integration limits for $R$ can be adjusted for $\xi$'s that
effectively vanish somewhere in the integration interval. Furthermore,
also the integration limits for $\bar{r}$ can be adjusted if
\mbox{$Q(\bar{r},\Delta)$} vanishes effectively somewhere within the
integration limits.

\section{Accuracy of Limber's equation}\label{accuracy}

\changed{The aim of this section is to assess the accuracy of Limber's
equation by comparing its predictions to the exact solution for the
angular correlation function.  For this purpose, different spatial
clustering models $\xi$ are used that are constant in time. Moreover,
the focus is on auto-correlations $\omega$, hence $p_1(r)=p_2(r)\equiv
p(r)$.}

In order to quantify the accuracy of Limber's approximation, the exact
solution Eq. \Ref{laststep} is solved by numerical integration, by
means of \Ref{omega0}, and compared to the result from Limber's
equation \Ref{limber}. In order to make a fair comparison, the same
small angle approximation as in \Ref{limber} is applied to the exact
equation, which as it is written in \Ref{omega0} does, of course, not
require small galaxy separations:
\begin{eqnarray}
  \frac{1}{1+\cos{\theta}}&\approx&\frac{1}{2}\;,\\
  \Delta&\approx&\frac{1}{2}\sqrt{R^2-\bar{r}^2\theta^2}\;,\\
  \bar{r}\sqrt{2(1-\cos{\theta})}&\approx&\bar{r}\theta\;.
\end{eqnarray}

\subsection{Model galaxy probability distribution}

The intention is to study the accuracy of Limber's equation for
relatively narrow distributions of galaxies,
\begin{equation}\label{pmodel}
  p(r)\propto r^2\exp{\left(-\frac{1}{2}\frac{(r-r_{\rm
          c})^2}{\sigma_{\rm p}^2}\right)}\;.
\end{equation}
Thus, galaxies selected by this filter are centred on $r_{\rm c}$ with
some scatter $\sigma_{\rm p}$. The prefactor $r^2$ accounts for the
apparent increase in the angular galaxy number density with distance
$r$ within the light-cone which only for moderately wide or wide
distributions becomes relevant. The prefactor has been added to make
the distribution more realistic.

The parameter $\sigma_{\rm p}$ in \Ref{pmodel} is not identical with
the $1\sigma$-scatter of the distribution, and $r_{\rm m}$ is not
identical with $r_{\rm c}$; this is only roughly correct for small
$x\equiv\sigma_{\rm p}/r_{\rm c}$. \changed{However, it can be shown
that the ratio \mbox{$y\equiv\sigma/r_{\rm m}$} is entirely a function
of $x$, namely approximately
\begin{equation}\label{ratio}
  y\approx
  x\frac{\sqrt{3x^4+1}}{3 x^2+1}\left[
  1+0.01x^2-\exp{(-1.29\,x^{-1.73})}\right]\;,
\end{equation}
with an accuracy of a few percent for \mbox{$0\le x\lesssim 2.5$}.
This relation is used in the following to find a $p(r)$ for given $y$
and $r_{\rm c}$.}

\subsection{Power-law spatial correlation functions}

\begin{figure}
  \begin{center}
    \psfig{file=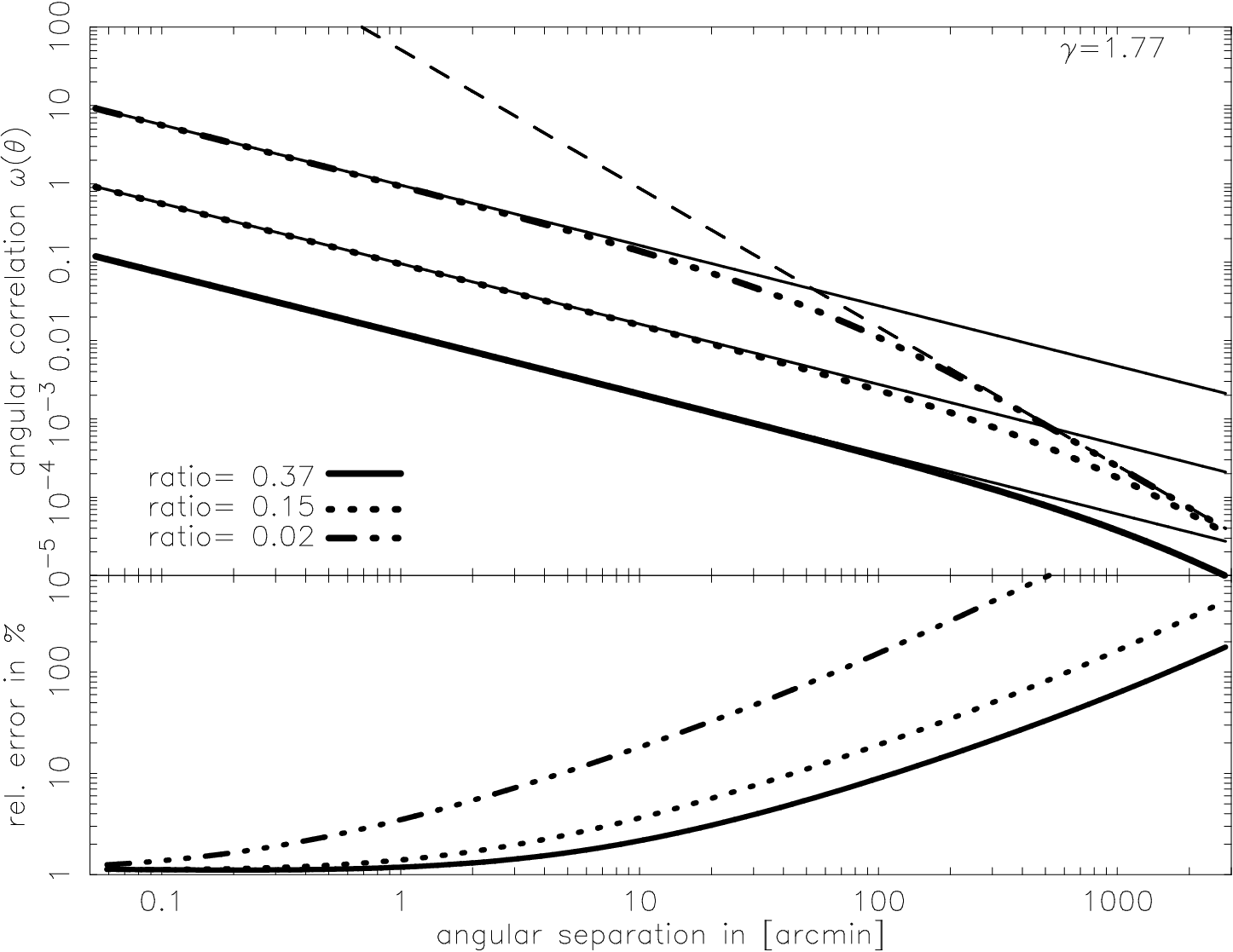,width=65mm,angle=-90}
  \end{center}
  \caption{\label{fig1a} Shown here is the angular correlation
    function $\omega$ for three different ratios
    \mbox{$\frac{\sigma}{r_{\rm m}}$} of the galaxy distribution, see
    Eq. \Ref{pmodel}, \mbox{$r_{\rm c}=2\,h^{-1}\,\rm Gpc$}. The
    spatial correlation, $\xi$, is a power-law with slope
    \mbox{$\gamma=1.77$} and clustering amplitude
    \mbox{$r_0=5.4\,h^{-1}\rm Mpc$} as roughly found for real
    galaxies. The thick lines are the numerical integrations of the
    exact equation, Eq.  \Ref{omega0}, while the thin solid lines are
    the predictions by Limber's equation, Eq.  \Ref{limberomega}.  The
    small angle approximations assumed in Limber's equation are also
    applied to the exact equation.  The relative difference between
    exact and Limber solution is plotted in the bottom small
    panel. The thin dashed line is the exact solution for an extremely
    peaked galaxy distribution, Eq.  \Ref{deltaomega2}.}
 \end{figure}

 The angular clustering, $\omega$, for three different galaxy
 distributions has been computed for the Figure \ref{fig1a}; $\xi$ is
 the same in all three cases. As expected, Limber's equation describes
 the exact solution quite well for small angular separations $\theta$
 but starts to deviate from the \mbox{$\gamma-1$} power-law for larger
 separations, becoming steeper and tending to look very much like the
 thin-layer solution \Ref{deltaomega2} (thin dashed line) which has
 the slope $\gamma$. Therefore, the exact angular clustering is a
 broken power-law if the spatial clustering is a power-law.  This
 behaviour is expected from the discussion in Appendix
 \ref{largertheta}.

 The break position is important for the Limber approximation: The
 closer one gets to the break position (and the farther beyond it),
 the more inaccurate is Limber's description. \changed{When $\xi$ can
 be assumed to be a power-law we can define a sensible position
 $\theta_{\rm break}$ for the break by working out where the Limber
 and the thin-layer approximation intersect. As shown in Appendix
 \ref{breakdetails} this is at:
 \begin{equation}\label{thetabreak2}
   \theta_{\rm break}\approx
   \frac{2\sqrt{3}}{\sqrt{\pi}}\frac{\Gamma(\gamma/2)}{\Gamma{(\gamma/2-1/2)}}\,
   \frac{\sigma}{r_{\rm m}}\;.
 \end{equation}}

 This relation reflects what has already been seen in the foregoing
 discussion. The break position is mainly determined by the ratio
 \mbox{$\sigma/r_{\rm m}$} -- but is also influenced by the slope
 $\gamma$ of $\xi$: the prefactor of $\sigma/r_{\rm m}$ in
 \Ref{thetabreak2} varies between $0.31$ and $1.18$ for
 \mbox{$\gamma=1.2,2.1$}, respectively: steeper $\xi$ are projected
 more accurately by Limber's equation than shallower $\xi$. 

  The relative error of Limber's approximation at the break position
  is already large, roughly $100\%$. Therefore, the validity regime of
  this approximation stops well below the break position. As a
  rule-of-thumb, the relative error is about $10\%$ at $\theta_{\rm
  break}/10$ which has been estimated from several plots like
  Figure \ref{fig1a}. Therefore, a $10\%$ error is reached at about
  \mbox{$\sim260\,\sigma/r_{\rm m}$ arcmin} for $\gamma\sim1.6$.

\subsection{More realistic galaxy clustering}\label{halomodel}

\changed{Although (spatial) galaxy clustering is empirically well
  described by a power-law on scales below \mbox{$r\sim 15\,\rm
  Mpc/h$}, $\xi(r)$ on large cosmological scales eventually adopts the
  shape of the dark matter clustering which does not have a constant
  slope $\gamma$ \citep[e.g.][]{2004ApJ...601....1W}. Therefore, the
  previous discussion is only applicable if the effective physical
  scales, that correspond to the angular separation interval under
  investigation, are below \mbox{$r\sim15\,\rm Mpc/h$}. In the
  opposite case, we expect that the accuracy of Limber's equation is
  not only a function of $\sigma/{r_{\rm m}}$ and an average of
  $\gamma$, but also a function of $r_{\rm m}$.}

\begin{figure}
  \begin{center}
    \psfig{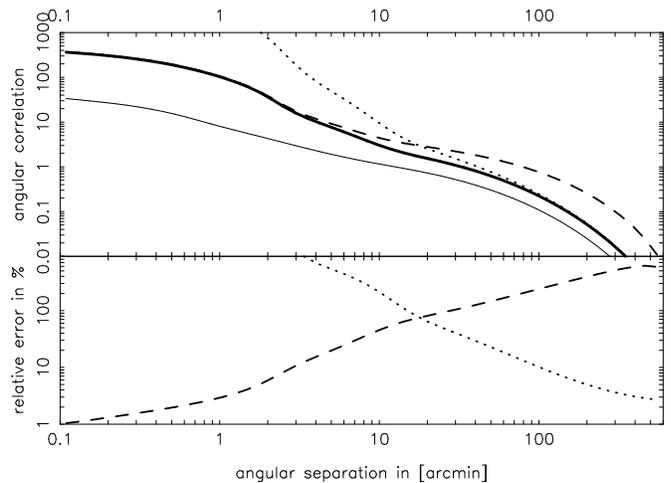}
  \end{center}
  \caption{\label{fig3} Angular clustering (arbitrary scaling) of the
  red galaxy sample (upper panel, thick solid line) centred on
  \mbox{$z=0.2$}, the relative width of the distribution in comoving
  distance is \mbox{$\sigma/r_{\rm m}=0.02$}. For comparison, also in
  the upper panel, the clustering of the blue population is plotted as
  well (thin solid line). The dashed and dotted line are the
  approximate solutions according to Limber's equation and the
  thin-layer approximation (upper panel). The lower panel depicts the
  relative error of the Limber (dashed) and thin-layer solution
  (dotted) for the red galaxies, respectively.}
 \end{figure}

\begin{figure}
  \begin{center}
    \psfig{file=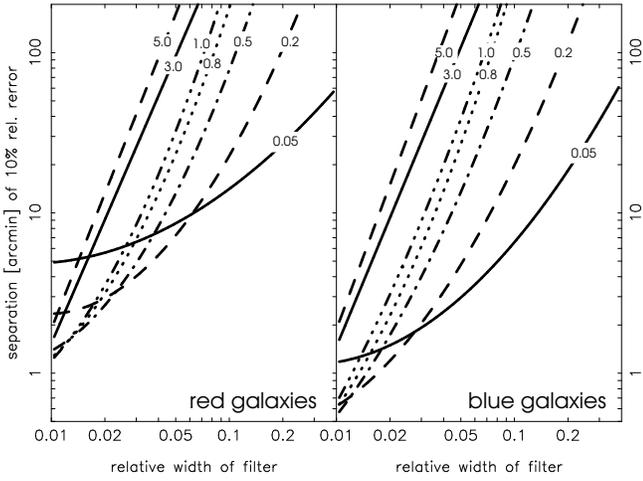,width=85mm,angle=0}
  \end{center}
  \caption{\label{fig4}Galaxy separation, as a function of
  \mbox{$\sigma/r_{\rm m}$} of the filter, at which Limber's equation
  becomes inaccurate by $10\%$ for red (left panel) and blue (right
  panel) galaxies. Different lines correspond to different $r_{\rm
  m}$; the corresponding redshifts are the line labels. Note that for
  estimating the relative error the \emph{local} correlation function
  of red and blue galaxies was simply moved to higher redshifts. The
  (flat) fiducial cosmology assumes \mbox{$\Omega_{\rm m}=0.3$}.}
 \end{figure}

\changed{To obtain more realistic $\xi(r)$, two representative models
  for galaxy clustering, based on a halo-model as in \citet[NFW
  parameters;][]{2003MNRAS.346..186M}, are used in this section: one
  $\xi$ is supposed to represent the clustering of galaxies that
  preferentially inhabit dark matter haloes of larger mass
  (\mbox{$M_{\rm halo}\ge10^{12.6}\,M_{\rm sun}$}; ``red galaxies''),
  another $\xi$ describes galaxies that also live in smaller mass
  haloes (\mbox{$M_{\rm halo}\ge10^{11}\,M_{\rm sun}$}; ``blue
  galaxies''). These two $\xi$'s are placed at different redshifts
  between \mbox{$0\le z\le5$}, are ``observed'' on the sky through
  filters $p(r)$ of varying widths $\sigma/r_{\rm m}$, and compared to
  Limber's equation in order to work out the relative error.  As to
  the accuracy of Limber's equation the amplitude of the correlation
  functions is irrelevant, it is only the shape that matters. For all
  redshifts considered the same $\xi$ is assumed, that is the
  halo-model prediction for \mbox{$z=0$}. To give an estimate of the
  relative error of narrow filters, this should be a fair
  assumption. See Figure \ref{fig3} for a particular example of the
  angular clustering.}

\changed{Figure \ref{fig4}. shows the estimated separation limits. The
  dependence of the filter on the centre, $r_{\rm m}$, however can
  clearly be seen, becoming weaker for larger $r_{\rm m}$. The
  difference between red and blue galaxies is most pronounced for
  small $r_{\rm m}$ but there are clear differences for
  \mbox{$\sigma/r_{\rm m}\lesssim0.02$}, even for large $r_{\rm
  m}$. Limber's prediction of the angular correlations of red galaxies
  is more accurate than the predictions for blue galaxies because on
  megaparsec-scales $\xi$ of red galaxies is steeper than $\xi$ of
  blue galaxies.}

\section{Summary and conclusions}\label{conclusions}

The Limber equation is an approximate solution for the angular
correlation function of objects, mainly applied in the context of
galaxies, for a given spatial correlation function. 

In this paper, it was shown that the assumptions made for Limber's
approximation become inaccurate \changed{beyond a certain angular
separation, which essentially depends on \mbox{$\sigma/r_{\rm m}$} and
the shape of the spatial correlation function}. It should be
emphasised that this inaccuracy has no relation to the flat-sky
approximation (small separations) that is commonly adopted when
working with Limber's equation.

Beyond what separation $\theta$ does Limber's approximation become
wrong by $10\%$?  For power-law $\xi$'s this separation depends
entirely on $\gamma$ and $\sigma/r_{\rm m}$ of the filter (selection
function) used. Spatial correlations with larger $\gamma$ are treated
more accurately by Limber's equation than $\xi$ with shallower
$\gamma$. As a rule-of-thumb, a systematic error of $10\%$ is reached
at about \mbox{$260\,\frac{\sigma}{r_{\rm m}}~\rm arcmin$} for
$\gamma\sim1.6$.

For more realistic $\xi$'s with a running slope $\gamma$, the
separation is also dependent on $r_{\rm m}$, the centre of the
filter. Even for relatively wide filters with \mbox{$\sigma/r_{\rm
m}\sim0.2$}, Limber's equation becomes inaccurate on the $10\%$ level
at a scale of a few degrees separation.  On the other hand, for narrow
filters of about \mbox{$\sigma/r_{\rm m}\lesssim0.02$}, a systematic
error of $10\%$ is already reached after $\sim10^\prime$ (or even a
few arcmin depending on $r_{\rm m}$).

\begin{figure}
  \begin{center}
    \psfig{file=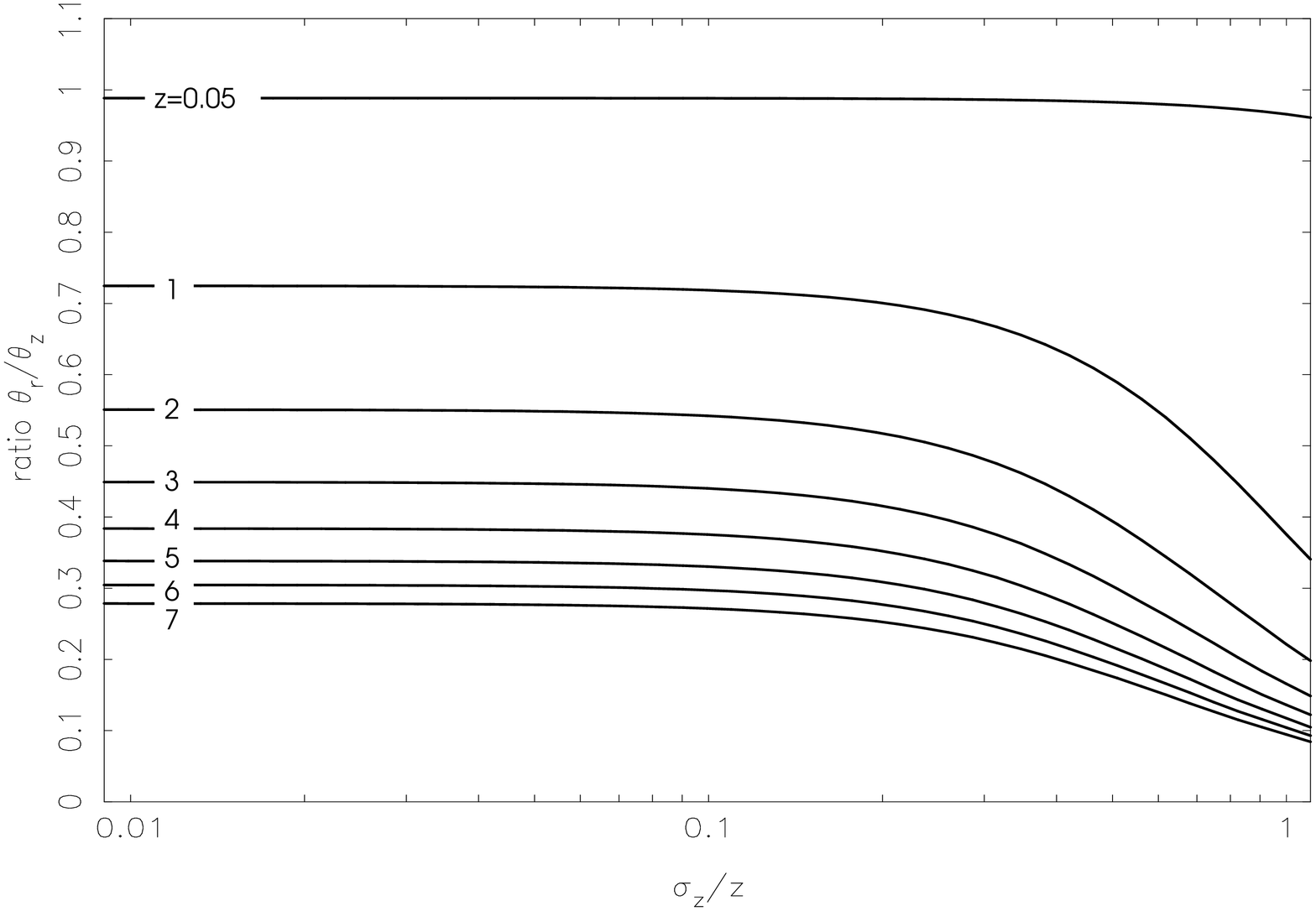,width=85mm,angle=0}
  \end{center}
  \caption{\label{fig2} \changed{Filter (galaxy distributions) $p(r)$
appear ``wider'' in redshift, \mbox{$p_{\rm z}(z)=p(r(z))\frac{\d
r}{\d z}$}, than in comoving distance, i.e. usually
\mbox{$(\theta_{\rm z}\equiv\sigma_{\rm z}/\bar{z})\ge(\theta_{\rm
r}\equiv\sigma_{\rm r}/r_{\rm m})$} where $\sigma_{\rm z}$ and
$\sigma_{\rm r}$ are the $1\sigma$-widths in redshift, and comoving
distance, respectively; $r_{\rm m}$ is the mean of the weight in
comoving distance, while $\bar{z}$ is the mean in redshift. To give an
estimate of the correction $\theta_{\rm r}/\theta_{\rm z}$ ($y$-axis) as
function of $\theta_{\rm z}$ ($x$-axis) $p_{\rm z}$'s are assumed that
are close to Gaussians. Each solid line corresponds to different means
$\bar{z}\in[0.05,7]$ (upper to lower line).}}
 \end{figure}

\begin{table}
  \center
  \begin{tabular}{lll}
    galaxy sample&$\bar{z}\pm\sigma_z$&$\sigma/r_{\rm m}$\\
    \hline\\
    \textbf{\cite{2007ApJ...654..138Q}}\\
    ~$K$-dropouts&$z=2\ldots3.5$&$\sim0.16$\\
    \textbf{\cite{lee06}}\\
    ~$U$-dropouts&$3.2\pm0.3$&$\sim0.04$\\
    ~$B_{435}$-dropouts&$3.8\pm0.3$&$\sim0.03$\\
    ~$V_{606}$-dropouts&$4.0\pm0.3$&$\sim0.02$\\
    \textbf{\cite{2006ApJ...648L...5O}}\\
    ~$i_{775}$-dropout&$6.0\pm0.5$&$\sim0.03$\\
    \textbf{\cite{adel05}}\\
    ~BM&$1.69\pm0.30$&$\sim0.12$\\
    ~BX&$2.24\pm0.37$&$\sim0.09$\\
    ~LBGs&$2.94\pm0.30$&$\sim0.05$\\
    \textbf{\cite{hildebrandt05}}\\
    ~LBGs&$2.96\pm0.24$&$\sim0.04$\\
    \textbf{\cite{ouchi04}}\\
    ~$BRi\rm-LBGs$&$4.0\pm0.5$&$\sim0.05$\\
    ~$Viz\rm-LBGs$&$4.7\pm0.5$&$\sim0.04$\\
    ~$Riz\rm-LBGs$&$4.9\pm0.3$&$\sim0.02$\\
    ~LAE&$4.86\pm0.03$&$\sim0.003$
  \end{tabular}
  \caption{\label{examples}\changed{List of typical values for the
  relative width of the survey selection function,
  \mbox{$\sigma/r_{\rm m}$} (comoving distance), in recent works using
  the dropout technique (LBGs: Lyman-break galaxies, LAE:
  Ly$\alpha$-emitter). Note that typically only separations less than
  about one arcmin are considered so that the accuracy of Limber's
  equation is, apart from the only LAE study in the list, no real
  issue.}}
\end{table}

Note that a filter can appear quite wide, i.e. \mbox{$\sigma_{\rm
z}/\bar{z}$} is large, as a function of redshift $z$ while it is
actually quite narrow as a function of comoving distance, the latter
being the relevant parameter for the accuracy of Limber's equation. In
particular, this means that applying Limber's equation to high
redshift galaxies may be a problem. The reader might find Figure
\ref{fig2} useful to quickly look up the approximate conversion
between $\sigma_{\rm z}/\bar{z}$ and $\sigma_{\rm r}/r_{\rm m}$.

Table \ref{examples} shows a small list of some recent studies of the
high-redshift universe using the dropout-technique. According to the
table the relative width of the selection filter frequently reaches
values near $0.03$. Therefore, Limber's equation should in these and
similar studies not be used when separations larger than
$\sim20^\prime$ are considered.

In cases where Limber's approximation is in doubt, that one uses the
exact Eq.  \Ref{omega0} that can easily be integrated numerically.

Limber's approximation is also used in gravitational lensing to
quantify correlations between the gravity-induced distortions of faint
galaxy images \citep[see e.g. ][]{bs01, kaiser98}. In this case, the
filter function relevant for lensing is the lensing efficiency that,
quite naturally, has a wide distribution; one can roughly estimate
\mbox{$\sigma/r_{\rm m}\approx0.22$} if the source galaxies have a
typical value of \mbox{$z_{\rm s}=1$}. The centre of this filter is at
about \mbox{$\bar{z}\approx0.5$}. These values will depend on the
distribution of source galaxies in redshift and the fiducial
cosmological model. Taking these values as a rough estimate of the
relative lensing filter width and centre, and assuming that the dark
matter clustering is somewhere between the clustering of red and blue
galaxies, one can infer from Figure \ref{fig4} that a two-point
\emph{auto-correlation} function of the convergence, based on Limber's
equation, is accurate to about $10\%$ for separations less than
several degrees; beyond that, an alternative description should be
used.
\begin{acknowledgements}
  I would like to thank Hendrik Hildebrandt and, in particular, Peter
  Schneider for carefully reading the manuscript, and their comments.
  This work was supported by the Deutsche Forschungsgemeinschaft (DFG)
  under the project SCHN 342/6--1.  Furthermore, I am grateful for the
  kind hospitality of the Karolinska Institute, Stockholm, and David
  Kitz Kr\"amer.
\end{acknowledgements}
\bibliographystyle{aa}
\bibliography{limber}
\appendix

\section{Details of Limber's approximation}\label{limberdetails}

In order to find an approximation of Eq. \Ref{laststep}, one
introduces for convenience new coordinates,
\begin{equation}
  \bar{r}\equiv\frac{r_1+r_2}{2}~~;~~\Delta r\equiv r_2-r_1\;,
\end{equation}
which are the mean radial comoving distance and difference of
radial distances, respectively, of a pair of galaxies.  This change of
coordinates renders \Ref{laststep} in the following manner:
\begin{eqnarray}\nonumber
&& \omega(\theta)=\\
&&\label{exactsol}
 \int_0^\infty\!\!\!\d\bar{r}\int_{-2\bar{r}}^{+2\bar{r}}\!\!\!\!\!\d\Delta r\,
 p_1\!\left(\bar{r}-\frac{\Delta r}{2}\right)\,
 p_2\!\left(\bar{r}+\frac{\Delta r}{2}\right)
 \xi(R^\prime,\bar{r})\;,
\end{eqnarray}
where 
\begin{equation}\label{rprime}
  R^\prime\equiv\sqrt{2\bar{r}^2(1-\cos{\theta})+\frac{\Delta r^2}{2}(1+\cos{\theta})}\;.
\end{equation}
This equation is simplified further by making the following
 approximations which are roughly satisfied for wide weight functions
 $p_{1,2}$, and for $\xi$ that fall off sufficiently fast over the typical
 width of $p_{1,2}$:
\begin{eqnarray}
  \label{diverge}
   p_1\!\left(\bar{r}-\frac{\Delta r}{2}\right)
   p_2\!\left(\bar{r}+\frac{\Delta r}{2}\right)&\approx&p_1(\bar{r})p_2(\bar{r})\;,\\
   \int_{-2\bar{r}}^{+2\bar{r}}\d \Delta r\,\xi(R^\prime,\bar{r})&\approx&
   \int_{-\infty}^{+\infty}\d \Delta r\,\xi(R^\prime,\bar{r})\;.
\end{eqnarray}

Especially the approximation \Ref{diverge} is characteristic for
Limber's equation. It is justified if the weight functions $p_{1,2}$
do not ``vary appreciably'' \citep[ p.43] {bs01} over the coherence
length of structures described by $\xi$ -- typically a few hundred Mpc
in the context of cosmological large-scale structure --, which means
we consider cases in which the coherence length is small compared to
the width of the weight functions $p_{1,2}$. In total, these
assumptions lead to the (relativistic) Limber equation
\citep{limber53,peebles80}:
\begin{equation}
  \omega(\theta)=
  \int_0^\infty\d\bar{r}\,p_1(\bar{r})p_2(\bar{r})
  \int_{-\infty}^{+\infty}\d{\Delta r}\,\xi(R,\bar{r})\;,
\end{equation}
where 
\begin{equation}
  R\equiv\sqrt{\bar{r}^2\theta^2+\Delta r^2}\;.
\end{equation}
For historical reasons, as a further approximation it is assumed in the
above equations that we are dealing with small angles of separation,
$\theta$, by which we can introduce:
\begin{equation}
  1+\cos{\theta}\approx2~~;~~
  1-\cos{\theta}\approx\frac{\theta^2}{2}\;.
\end{equation}
These two approximations are accurate to about $10\%$ for angles
smaller than \mbox{$\theta\lesssim40^\circ$} which covers the typical
range of investigated separations. Usually, when employing Limber's
equation this approximation is automatically used.

\section{Approximation for larger separations}\label{largertheta}

It will be shown here that the thin-layer solution of the Sect.
\ref{thinlayer} is in fact asymptotically approached if the separation
$\theta$ only gets large enough.

To see why the exact solution behaves like this we have to go back to
Eq. \Ref{exactsol} and in particular \Ref{rprime}. From that we can
work out what happens for larger $\theta$. Note that the following
focuses on auto-correlations. The $\Delta r$-term in \Ref{rprime} can
be neglected compared to the $\bar{r}$-term if
\begin{eqnarray}
  2\bar{r}^2(1-\cos{\theta})&\gg&\frac{\Delta r^2}{2}(1+\cos{\theta})\\\label{tancon}
  \Longleftrightarrow \frac{\Delta r}{2\bar{r}}\tan^{-1}{\left(\frac{\theta}{2}\right)}&\ll&1\;.
\end{eqnarray}
Let us say the distribution $p$ of galaxies has a characteristic width
(variance) of $\sigma$ and a mean of $r_{\rm m}$. Pairs of galaxies
from this distribution will typically have
\mbox{$\bar{r}=\ave{(r_1+r_2)/2}\approx r_{\rm m}$} and for their
mutual distance \mbox{$\Delta
  r\approx\sqrt{\ave{(r_1-r_2)^2}}\approx\sqrt{2}\sigma$}.  Therefore,
the condition \Ref{tancon} will be given for $\theta$ large enough to
fulfil
\begin{equation}\label{largetheta2}
  \tan{\left(\frac{\theta}{2}\right)}\gg\frac{\sigma}{\sqrt{2}r_{\rm m}}\;,
\end{equation}
or for small $\theta$ ($10\%$ accuracy for
\mbox{$\theta\lesssim60^\circ$}) approximately
\begin{equation}
  \theta\gg\sqrt{2}\frac{\sigma}{r_{\rm m}}\;.
\end{equation}
In this regime, where the $\Delta r$-term in $R^\prime$ is negligible,
Eq. \Ref{exactsol} simplifies to
\begin{equation}\label{largetheta}
  \omega(\theta)=\int_0^{+\infty}\d\bar{r}
  \,F(\bar{r})\,\xi\!\left(\bar{r}\,\sqrt{2}\sqrt{1-\cos{\theta}},\bar{r}\right)\;,
\end{equation}
which is a family of thin-layer solutions \Ref{deltaomega2} weighted
averaged with the kernel
\begin{equation}
  F(\bar{r})\equiv\int_{-2\bar{r}}^{+2\bar{r}}\d\Delta r\,
  p\!\left(\bar{r}+\frac{\Delta r}{2}\right)\,
  p\!\left(\bar{r}-\frac{\Delta r}{2}\right)\;.
\end{equation}
 Already for moderately wide $p(r)$, $F(r)$ is relatively peaked, so
 that for large enough $\theta$, Eq. \Ref{largetheta} is close to the
 solution \Ref{deltaomega2} with $r_{\rm c}\approx r_{\rm m}$. Note
 that $F(r)$ is normalised to one.

\section{Break position}\label{breakdetails}

Limber's equation is an accurate description for small $\theta$ and
the ``thin-layer'' solution, Eq.  \Ref{deltaomega2}, is an approximate
solution for larger $\theta$. For power-law like spatial correlation
functions, \mbox{$\xi\propto r^{-\gamma}$}, a sensible definition for
the position of the break is at the angle $\theta_{\rm break}$ (in
RAD) where both approximations intersect:
\begin{equation}\label{characteristic}
  A_\omega\,\theta_{\rm
  break}^{1-\gamma}= \xi(\sqrt{2}\sqrt{1-\cos{\theta_{\rm break}}}r_{\rm m})\;.
\end{equation}
 Using \Ref{powerlawxi} and \Ref{aomega} one obtains after some
 algebra:
 \begin{eqnarray}\nonumber
  && \theta_{\rm break}^{1-\gamma}\left(\sqrt{2}\sqrt{1-\cos{\theta_{\rm
	 break}}}\right)^\gamma=\\\label{thetabreak}
   &&   \frac{\Gamma(\gamma/2)}{\sqrt{\pi}\,\Gamma(\gamma/2-1/2)}
   \left[r_{\rm m}^\gamma\int_0^\infty\d\bar{r}\,p^2(\bar{r})
     \bar{r}^{1-\gamma}\right]^{-1}\;.
 \end{eqnarray}
 For \mbox{$\gamma\in[1,2]$} the left-hand side of the previous
 equation is for \mbox{$\theta_{\rm break}\lesssim60^\circ$} within
 $10\%$ accuracy just
 \begin{equation}\label{approx2}
   \theta_{\rm break}^{1-\gamma}\left(\sqrt{2}\sqrt{1-\cos{\theta_{\rm
	 break}}}\right)^\gamma\approx\theta_{\rm break}\;.
 \end{equation}
 Thus, this is a reasonable approximation to use.

 The relation \Ref{thetabreak} is useful because it allows one to
 estimate the expected position of the break in the power-law
 behaviour of $\omega$.  As the accuracy of Limber's equation is
 primarily an issue for narrow $p$ we can derive from \Ref{thetabreak}
 a rule-of-thumb for the break position. For narrow $p$, it is
 sensible to assume a top-hat function as in Eq. \Ref{tophat}. Then,
 employing \Ref{approx}, \Ref{approx2} and \Ref{thetabreak} one gets:
\begin{equation}
  \theta_{\rm break}=
  \frac{2\sqrt{3}}{\sqrt{\pi}}\frac{\Gamma(\gamma/2)}{\Gamma{(\gamma/2-1/2)}}\,
  \frac{\sigma}{r_{\rm m}}\;.
\end{equation}
Note that the variance of the top-hat is \mbox{$\sigma=\Delta
  r/\sqrt{3}$, where $\Delta r$ is the width of the top-hat}. 

\end{document}